\documentstyle[emulateapj]{article}   

\lefthead{S.\ Cassisi et al.}   
   
\righthead{He-flash induced mixing in Pop~II stars}   
   
\begin{document}   
\title{First full evolutionary computation of the He-flash induced mixing   
in Population~II stars.}   
   
\author{Santi Cassisi\altaffilmark{1,3}, Helmut  
Schlattl\altaffilmark{2,3}, Maurizio Salaris\altaffilmark{2,3} and  
Achim Weiss\altaffilmark{3}}   
   
\altaffiltext{1}{INAF - Osservatorio Astronomico di Collurania, Via M.\ Maggini,    
I-64100 Teramo, Italy; cassisi@te.astro.it}   
   
\altaffiltext{2}{Astrophysics Research Institute, Liverpool John Moores University,   
Twelve Quays House, Egerton Wharf, Birkenhead CH41 1LD, UK;hs,ms@astro.livjm.ac.uk}   
   
\altaffiltext{3}{Max-Planck-Institut f{\"u}r Astrophysik, Karl-Schwarzschild-Strasse 1,   
D-85748 Garching, Germany; aweiss@mpa-garching.mpg.de }   
   
\begin{abstract}   
The core helium-flash in low-mass stars with extreme mass loss occurs   
after the tip of the RGB, when the H-rich envelope is very thin. The   
low efficiency of the H-shell source enables the He-flash driven convective zone   
to penetrate H-rich layers and trigger a thermonuclear runaway,   
resulting in a subsequent surface enrichment with He and C. In   
this work we present the first full computations of  
Population~II low-mass stellar models  
through this phase. Models experiencing this  
dredge-up event are significantly hotter than   
their counterparts with H-rich envelopes, which makes them promising   
candidates for explaining the existence of  
stars observed beyond the canonical blue end of the   
horizontal branch (``blue hook stars''). Moreover, this temperature   
difference could explain the observed gap in $M_V$ between extreme   
blue horizontal-branch and blue hook stars. A first  
comparison with spectroscopic observations of blue hook stars in  
the globular cluster $\omega$~Cen is also presented.  
\end{abstract}   
   
\keywords{stars: abundances --- stars: evolution --- stars: horizontal 
branch --- stars: late-type}
   
\section{Introduction}   
   
The Horizontal Branch (HB) in the Color-Magnitude Diagram (CMD) of a   
number of Galactic Globular Clusters (GGCs), like e.g.\ NGC~6752,    
shows extended blue tails   
populated by so called Extreme HB (EHB) stars,   
characterized by extremely high effective temperatures.   
The existence of these   
EHB stars is hard to interpret in terms of canonical stellar   
evolution, where the He-flash takes place at the tip of the
Red Giant Branch (RGB)\footnote{In this sense we understand the canonical
HB to consist of stars igniting He at the tip of the RGB.}. None of those stars
would have a hot enough 
HB locus to explain the   
extended blue tail of, e.g., NGC~6752 (see, e.g., Moehler et al.~2000).   
Castellani \& Castellani~(1993) have suggested the   
so-called ``delayed helium-flash'' (``delayed HEF'') scenario to explain the   
existence of EHB stars. This scenario    
envisages that, as a consequence of an high   
mass-loss efficiency during the Red Giant Branch (RGB)    
evolution (due to enhanced stellar wind and/or dynamical interactions with other   
stars in the dense cluster core), a star can lose so much envelope   
mass that it    
fails to ignite the HEF at the tip of the RGB, thus    
evolving toward the White Dwarf (WD) cooling sequence with an   
electron-degenerate helium core.   
Depending on the amount of the residual H-rich envelope mass, the star will   
still ignite He-burning either     
at the bright end of the WD cooling sequence ("early" hot flasher, hereafter   
EHF), or along the WD cooling sequence ("late"    
hot flasher, hereafter LHF).   
After the He-flash these stars will settle on the Zero Age HB (ZAHB) 
and with their strongly reduced envelope mass they are much hotter
than their counterparts on the canonical ZAHB.
   
Recent far-$UV$ photometric observations of the GGC   
$\omega$~Cen (see e.g. D'Cruz et al.~2002) and NGC2808   
(Brown et al. 2001, B01) have disclosed the existence   
of a spread in the $UV$ brightness at the   
hot end of the observed HB stars, larger than the expected photometric   
errors, and with a sizable number of stars appearing to form a hook-like   
feature. These "blue hook" stars are located even beyond the hot end   
of EHB objects like the ones observed in NGC~6752 (Moehler et al.~2002, M02), 
and are fainter than the bluest end of the canonical ZAHB locus by up
to about 0.7\,mag in the $UV$.
The evolution of EHFs and their possible connection with blue hook   
stars have been investigated by D'Cruz et al. (2000), but the   
progeny of such EHFs is not hot enough to explain the existence of    
"blue hook" stars.

A more promising working scenario has been suggested by Sweigart (1997, S97)
based on the evolution of LHFs. He
finds that the convection zone produced by the late HEF is   
able to penetrate into the H-rich envelope, thereby mixing H into    
the hot He-burning interior (He-flash mixing, hereafter HEFM)   
where it is  burned rapidly.    
A consequent dredge-up of matter which has    
been processed via both H- and He-burning enriches   
the outer envelope by He   
and some carbon and nitrogen.    
According to S97 these abundance changes   
cause a discontinuous increase of the HB effective temperature   
at the transition between unmixed and mixed models, producing   
a gap at the hot end of the HB stellar distribution    
as indeed observed by Bedin et al. (2000) in NGC2808. At the same time the     
changes of the surface chemical composition induced by the HEFM    
modify the emergent spectral energy distribution, and may explain   
the fact that these stars appear as subluminous objects in the $UV$
CMD.    

This scenario has been examined in more detail by B01,
but their computations were stopped at the onset of the HEFM,
due to the numerical difficulty of following the development of this process and the   
simultaneous nucleosynthesis within the convection zone.  
B01 could therefore only estimate the surface chemical abundances after  
the HEFM episode, and the resulting ZAHB effective temperature.    
In this paper we present the first full computation of   
LHF star evolution, following the development and outcome   
of the HEF and HEFM; in this way we are able to consistently predict the effect of   
the HEFM process on the chemical stratification of the stellar   
envelope, thus eliminating the need to assume an ad-hoc chemical distribution for the   
post-HEFM phase.   
 
In \S\ref{comp} we briefly introduce our computational code and   
present the evolutionary sequences for LHF stars. The resulting model   
properties are compared with the models and estimates of B01.   
In the final section our results are briefly compared with  
relevant observations.

\section{Evolution of LHF stars}\label{comp}   
 
A consistent calculation of the HEFM process is challenging,   
because the timescale for proton-capture reactions within the   
flash-convection zone is comparable to the convective mixing   
timescale; therefore the latter has to be   
treated time-dependently and   
solved simultaneously with the nuclear network.   
Due to the evident numerical difficulties in the only available
computation of a     
low-mass star experiencing HEFM performed by S97, the energy    
contribution provided by the H-burning was not accounted for. This is clearly a too crude approximation   
since it causes an underestimate of the extent of the mixed   
region and, thus, of the final effects on the chemical composition of the outer layers.   
   
In order to perform the first full evolutionary calculations of stars   
undergoing HEFM,    
we adopt an evolutionary code (Schlattl et al.~2001, 2002, hereafter   
S01, S02) in which the equations for convective mixing    
and nuclear burning are solved in one common scheme, and which    
accounts for the simultaneous burning of both H and He.    
The time-dependent mixing is treated   
as a fast diffusive process, where the diffusion coefficient is   
proportional to the     
convective velocity derived from a convection theory (S01).    
The reliability and accuracy of this approach has been already   
discussed by S01 and S02, in their investigation   
of HEFM  in extremely metal-poor stars   
during the core He-flash at the tip of the RGB.    
 
We have computed several sequences of evolutionary models starting from   
the same 0.862\,$M_\odot$ progenitor, with initial metallicity   
$Z=0.0015$ and helium abundance $Y=0.23$.   
This choice is the same made by B01, and    
provides an age for the models at the RGB tip of about 12.6\,Gyr,    
in agreement with current estimates of GGC ages (e.g., Salaris \& Weiss~2002).   
The sequences are characterized by different amount of mass loss along the RGB,   
parameterized according to the standard Reimers~(1975) relation; we   
will denote the   
different sequences according to the value adopted for the free   
mass-loss parameter $\eta$.    
We have considered values of $\eta$  in the range between 0.0 and 0.65, in order to have models   
igniting the HEF at the RGB tip, as EHFs or as LHFs.   
   
Figure~1 shows the evolution in the H-R diagram of models    
computed with different values of $\eta$, which   
experience the HEF in different evolutionary phases.    
We have found that the largest $\eta$    
leading to the HEF at the RGB tip is 0.55, whereas   
a value of about 0.58 provides    
a mass-loss efficiency corresponding to   
EHF stars. These two se- 
\centerline{\null} 
\vskip1.05truein   
\includegraphics{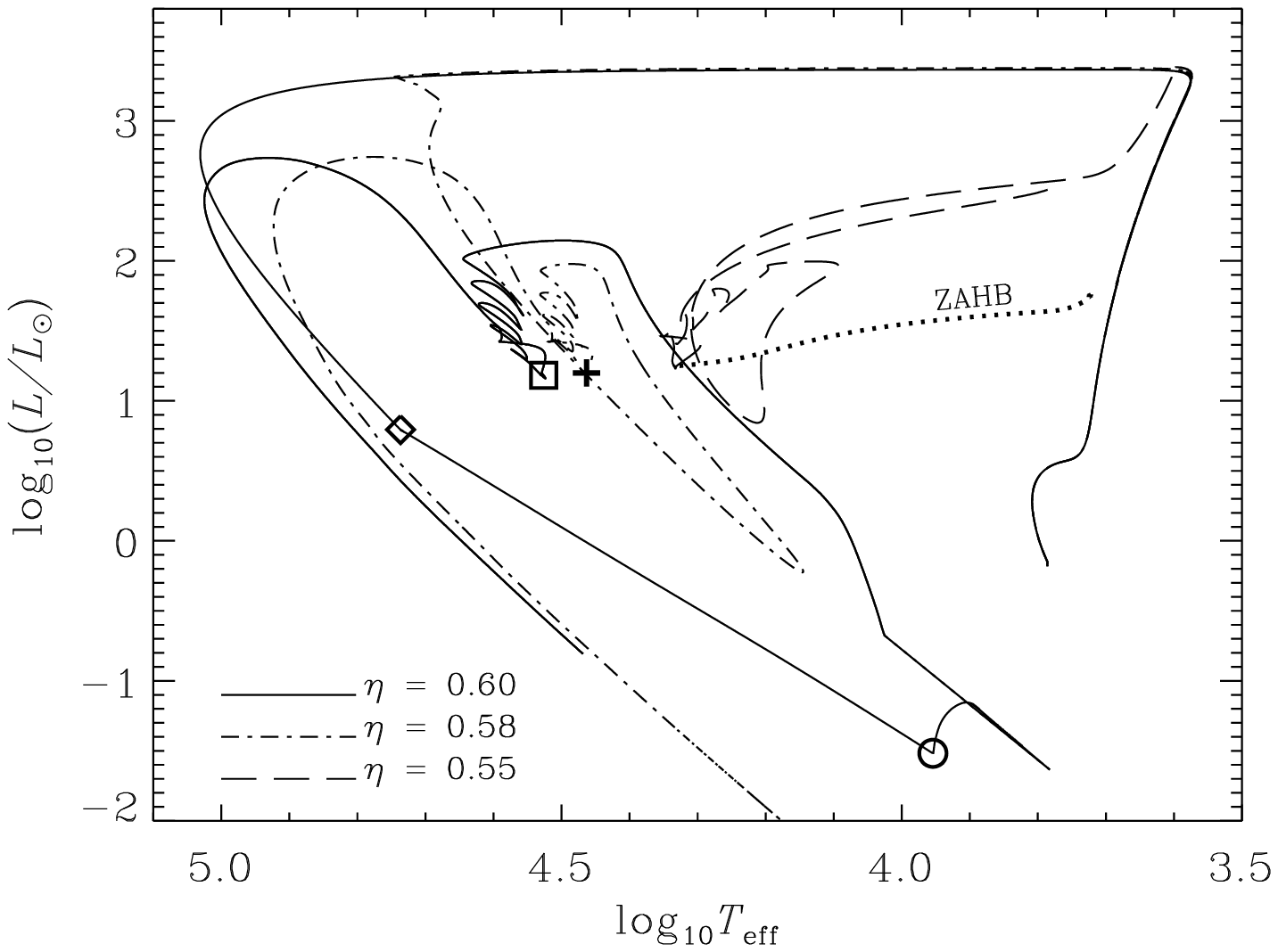}   
\vskip1.34truein   
\figcaption{The H-R diagram of some evolutionary models computed
accounting for different values (as labeled) of the Reimers mass-loss
parameter. The canonical ZAHB is also shown (dotted line) together
with the evolution along the Main Sequence and RGB of the
progenitor. The ZAHB location of the sequence with $\eta=0.58$ is
marked by `$+$', while the onset of the HEF, the H-mixing event, and
the consequent ZAHB location of the $\eta=0.60$ sequence are denoted
by `$\Diamond$', `{$\bigcirc$}', and `$\Box$', respectively.   
\label{fig1}}   
\vskip0.15truein 
\noindent quences have the same He-core mass
(0.4930$\,M_\odot$)     
at the HEF but very different envelope masses of 0.0192$\,M_\odot$   
and $0.0012\,M_\odot$, respectively, which leads to a difference of
$\sim 7\,500\;$K in 
the ZAHB effective temperature of these models    
(i.e. $\sim 21\,500\;$K     
against $\sim 29\,000\;$K, respectively).
These two models do not experience HEFM ---   
thus maintain H-rich envelopes --- and   
share the same structural properties during the major HEF as   
expected on the basis of the results obtained by Castellani \&
Castellani~(1993) and B01.    

The situation changes drastically when a slightly higher value of   
$\eta$ is used.   
The sequence computed with $\eta=0.60$ 
ignites  
the HEF along the WD cooling sequence; in this case   
the events following the core HEF are completely   
different from the evolution of models computed with a   
lower $\eta$.   
The HEF ignition is located 0.141$\,M_\odot$ off center\footnote{We 
consider the HEF to start when the He-burning luminosity is     
equal to $100\,L_\odot$, as in Sweigart \& Gross~(1978).},    
when $\log{(L/L_\odot)}=0.79$; in this phase the luminosity   
produced by the H-burning shell is $\sim0.35\,L_\odot$ and the He-core mass is equal to   
0.4907$\,M_\odot$, i.e., $0.0023\,M_\odot$ lower than in the model   
computed with $\eta$=0.58. The envelope mass is equal to
$5.5\times10^{-4}\,M_\odot$, about half the value of the model with
$\eta=0.58$.    
   
Soon after the HEF ignition    
the surface luminosity drops and the peak of the flash luminosity   
($L_{\rm He}=3.7\times10^{10}\,L_\odot$) occurs when the    
surface luminosity is equal to $\log{(L/L_\odot)}=0.36$,   
about $1\,000$ years after the onset of the HEF. The expansion of the core   
during the flash episode cools down the   
surrounding layers, and when the HEF is at its peak the energy   
contribution provided by the H-burning shell drops to   
$L_{\rm H}\approx4.9\times10^{-3}\,L_\odot$.    
Due to the huge energy flux caused by the HEF a   
convective region develops above the point where He ignites. This convection zone reaches the   
H-rich envelope and mixes protons from the envelope into the hot   
He-burning interior when the stellar   
luminosity is equal to $\log{(L/L_\odot)}=-1.2$ and the energy
provided by H- and He-burning are $2.7\times10^{-6}\,L_\odot$ and $1.1\times10^{7}\,L_\odot$, respectively.    
This HEFM
\centerline{\null} 
\vskip1.38truein   
\includegraphics{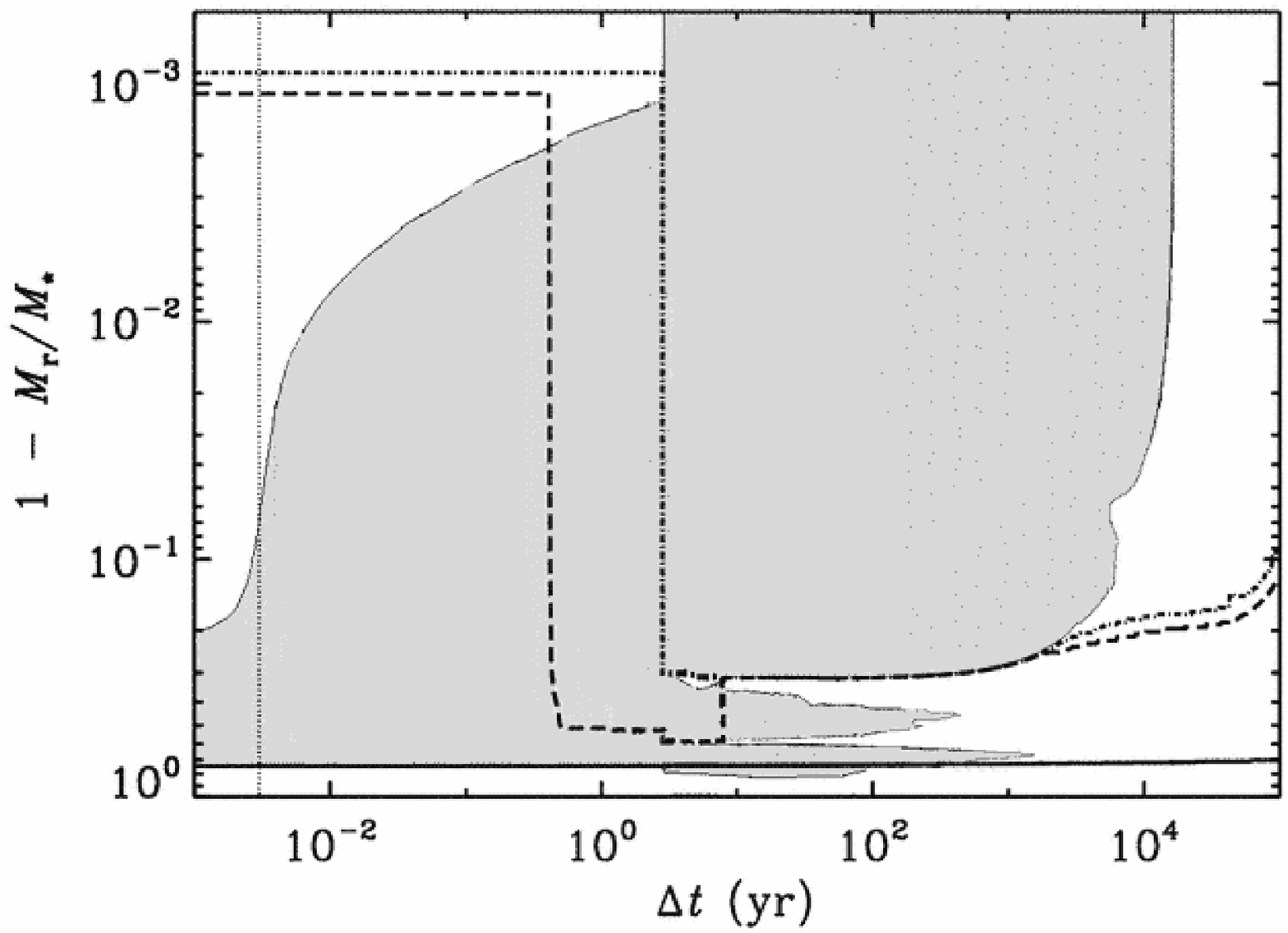}   
\vskip1.0truein 
\figcaption{Evolution from the beginning of the  
HEF of the different convective zones,   
indicated by shaded areas, in our model with $\eta=0.60$. The solid, dashed,
and dot-dashed line denote the mass location of maximum    
energy release by He-burning, CNO cycle, and $p$-$p$ chain, respectively.   
The vertical dotted line marks the moment of the HEF peak luminosity; 
the zero point in time has been chosen to be one day before this moment. 
\label{fig2}}   
\vskip0.2truein   
\noindent     
 event occurs about 5 months after the HEF was at maximum.   
   
The evolution and the main properties of this LHF model up to this point   
are in good agreement with the results obtained by B01 for the corresponding   
model with the smallest mass-loss efficiency necessary to undergo HEFM.    
B01's values for the He-core mass and envelope mass are   
0.491$\,M_\odot$ and $6.0\times10^{-4}\,M_\odot$, respectively, very   
similar to our results. However, the minimum    
$\eta$ necessary to have a LHF with HEFM is equal to 0.818 in B01   
models, while we need a much lower value of 0.60, which means that our   
models lose more mass than the ones of B01.   
Our models are on average   
cooler than theirs during the HB phase (e.g., the ZAHB effective   
temperature of our hottest model not experiencing HEFM is lower than   
in B01) and therefore they are possibly also cooler along the RGB   
where large part of the mass loss happens. Since, according to the   
Reimers relation cooler models tend to lose more mass at a fixed   
value of $\eta$, this would explain the smaller value of $\eta$ needed   
in our models. The fact that our hottest model without HEFM is 
cooler by about $1\,500\;$K than the one in B01, cannot be explained 
by our coarser $\eta$ resolution, as, e.g., models with $\eta=0.58$ and $0.59$
differ only by about 500\,K in their ZAHB location~(Table~\ref{zahbdat}).
   
Whereas B01 stopped their computation at the beginning of   
the H ingestion by the He-flash driven convection zone, we are able to   
follow consistently the evolution through this event, and thus    
obtain reliable estimates of the variation of the relevant surface   
chemical abundances. This is important not only in order to   
predict the spectral energy distribution of these stars,   
but also to provide quantitative estimates for comparison with  
spectroscopic observations of "blue hook" stars.   

The evolution of the convective zones in the   
$\eta=0.60$ sequence after the onset of the HEF, as well   
as the location of the layers where the H- and He-burning energy   
release are at maximum, are shown in Figure~2.   
As soon as H is carried into the interior,    
it starts burning through the CNO cycle   
at an extremely large rate, because of the high temperatures in these   
layers and the large abundance of    
$^{12}{\rm C}$ nuclei produced by the 3$\alpha$ reactions; at the same time   
the upper boundary of the flash-induced convective region moves   
further outward, eventually reaching the stellar surface.   
The maximum energy released by H burning is equal to   
$8.2\times10^{9}\,L_\odot$; 
the stellar luminosity then is equal to $\log{(L/L_\odot)}=-1.37$.   
 
The main consequence of the strong H-burning is   
that the flash-induced convective zone splits initially into
two and then even into three separate convective regions because of temporarily three 
active burning shells, i.e., the He-burning shell and two shells
related to H-burning through the $p$-$p$ chain and the CNO cycle, respectively
(Fig.~\ref{fig2}).  
We note that this process is similar to the one    
occurring in low-mass, extremely metal-poor stars at the RGB tip, see   
S01 and also B01. Because of the extremely thin envelope mass, the   
star does not expand to a RGB configuration, despite the large   
additional amount of energy injected into the envelope by the H-burning    
shell.    
The total amount of H burned during the HEFM episode is about   
$3.7\times10^{-4}M_\odot$ and the final mass fraction of H in the   
envelope is    
$4\times10^{-4}$, which is very low but still non-negligible (see   
\S\ref{final}).    
A large amount of matter   
processed by both H- and He-burning is dredged to the surface,   
changing its chemical composition.    
In particular, the final total metallicity $Z$ is equal to 0.0365; the   
metal distribution is strongly C-enhanced, the carbon mass fraction being  
equal to $0.029$, while the surface abundances of oxygen and nitrogen are    
$3.5\times10^{-5}$ and $0.007$, respectively. In addition,   
the surface He abundance increases to $Y=0.96$.   
 
After the envelope H is strongly reduced and the electron degeneracy is   
lifted in the core region close to the HEF ignition, the three convective   
shells disappear. Like in  standard He-flash models the He-burning   
shell slowly moves toward the center via a few additional   
HEFs (no HEFM ensues), removing    
the electron degeneracy in the whole He-core.   
Finally, the star settles on its ZAHB location at $\log(L/L_\odot)=1.18$   
and $T_{\rm eff} = 33\,600\;{\rm K}$ (see Fig.~1); the corresponding    
bolometric magnitude is about 0.14\,mag higher
than for the hottest canonical ZAHB model,   
due to the slightly smaller He-core mass at the HEF.  
(A summary of the ZAHB locations of the models with different 
mass-loss parameter $\eta$ is provided 
in Table \ref{zahbdat}.) 

The HB progeny of the $\eta=0.59$ model, i.e., the least massive one   
which does not experience the HEFM process, shows a ZAHB effective   
temperature equal to about    
$30\,000\;{\rm K}$ (close to ``+'' in Fig.~1). The    
discontinuity of the ZAHB effective temperature between    
models with and without HEFM is therefore $\sim3\,600\;{\rm K}$.   
This value is however only a lower limit to the real    
discontinuity;    
in fact, although radiative opacities for the correct He and H
abun\-dances have been used, the metal distribution was taken to be   
scaled solar, not C-enriched as would be   
appropriate for the post HEFM objects. A comparison of OPAL (see,
\centerline{\null}
\vskip-.07truein 
\renewcommand{\arraystretch}{1.25} 
\tabcaption{\centerline{ZAHB location of selected models\label{zahbdat}}}
\begin{tabular}{lccccc}\hline\hline 
$\eta$            & 0 & 0.55$^{\rm a}$ & 0.58 & 0.59 & 0.60 \\ \hline 
$\log(L/L_\odot)$ & 1.76 & 1.24 & 1.20 & 1.19 & 1.18 \\ 
$\log T_{\rm eff}$ & 3.722 & 4.332 & 4.463 & 4.474 & 4.526 \\ \hline 
\multicolumn{6}{l}{\footnotesize $^{\rm a}$~Hot end of canonical HB} 
\end{tabular} 
\noindent  e.g.,   
Iglesias \& Rogers~1996) scaled-solar opacity tables   
with C-enriched ones for $Y$ and $Z$ values nearly equal to   
the actual surface composition of our $\eta=0.60$ model reveals that   
the radiative opacities have been overestimated  
by up to 50\%.  

By crudely applying this correction to the opacity in a   
post-HEFM model before settling on its ZAHB, we estimate that the ZAHB 
location    
of our C-enriched models should be about $1\,500\;{\rm K}$ hotter. This    
increases the $T_{\rm eff}$-discontinuity between models with and   
without HEFM to $\sim5\,100\;{\rm K}$.  
Although B01 could not obtain the final surface chemical composition of   
their LHFs from full evolutionary computations, 
their post-HEFM ZAHB models with He-C mixture assume    
$Y=0.96$ and a carbon mass fraction of 0.04, both values being very   
similar to the outcome of our detailed computation.   
Employing the appropriate opacities B01 determined a discontinuity in
the ZAHB $T_{\rm eff}$ distribution    
of $\approx6\,000\;{\rm K}$, similar to the value we would obtain
in case of using opacities with the correct metal distribution. 

\section{Final remarks}\label{final}   
   
In the previous section we have shown that our full computations through    
the HEFM phase do support the evolutionary scenario for blue hook   
stars outlined by B01.    
The most important changes induced by the HEFM process are related to  
the surface chemical   
composition, which is   
modified in a drastic way with respect to the initial pattern   
due to the mixing of matter processed by both H- and He-burning.    
In order to compare theory with observations one needs   
spectroscopic analyses of samples of blue hook stars in GGCs.   
   
Very recently, a spectroscopic analysis of a sample of   
blue hook stars in $\omega$~Cen has been published by M02.   
On the basis of medium resolution spectra M02 found that the blue hook  
stars in $\omega$~Cen   
are significantly   
hotter ($T_{\rm eff}\ge35,000\;{\rm K}$) than EHB stars, have on   
average a solar He-abundance    
with four objects characterized by $Y\ge0.7$, and there are indications that   
their outer layers are C-enhanced, albeit by possibly much less than 
the value predicted by our models (analogous discrepancy does exist for 
the S02 computations of the HEFM in extremely metal-poor stars).  
These results provide some support   
for the HEFM scenario as a plausible explanation for the existence of   
blue hook stars. 
Another important result obtained by M02   
is the clear evidence that blue hook stars in $\omega$~Cen retain a  
substantial fraction of their initial H abundance, which is
barely reproduced by our models. For instance, in the $\eta=0.60$ model
we obtain    
$\log(n_{\rm He}/n_{\rm H})\approx2.8$ whereas the larger value   
measured by M02 is of the order of $+0.94\pm0.14$, all other values   
being negative.  

A possible cause of this discrepancy is   
the outward diffusion of H and the gravitational   
settling of He in the atmospheres of blue hook stars, as already   
suggested by M02, a process which is not included in our computations.   
One has also    
to bear in mind that the residual amount of H left at the surface after the HEFM process 
is strongly dependent on the mixing efficiency adopted in the   
evolutionary computations. A smaller efficiency would cause H to be   
burned at a smaller depth within the star, which would result in a 
thinner  
convective region   
above the H-burning shell, and thus a reduced H-dilution when this   
zone reaches the surface. Moreover, with the temperature being lower,   
H would be consumed at a smaller rate.   
Even if we have not performed a detailed   
investigation of the effect of a reduced mixing efficiency on our   
models, we can use our computations of the evolution of   
extremely metal-poor low-mass stars during the HEF (see S02)
to predict that a reduction of the mixing efficiency by a factor    
$2\times10^{4}$ should increase the residual amount of H by an order   
of magnitude. This would be in better agreement with the smaller value   
of the ratio $\log(n_{\rm He}/n_{\rm H})$ measured  
in one blue hook star by M02, but still   
too large to explain    
the higher H-abundances observed in all other blue hook stars of the   
M02 sample. It is also worth mentioning that the observed     
$T_{\rm eff}$ values are generally higher than the values attained by   
our LHF models.
   
Before a definitive assessment of the adequacy of the HEFM scenario   
in explaining the HB blue hook stars is possible, a large   
observational and theoretical    
effort is necessary; in particular, one needs to obtain higher quality   
spectra for a larger sample of GGC blue hook stars   
to investigate in more detail the evolutionary   
consequences of the HEFM process on a larger set of models and 
to study the dependence of the HEFM   
surface chemical composition changes on the numerical   
and physical assumptions, on which the computation of the
evolution  through this phase are based. This last topic   
will be addressed in a forthcoming paper.

\acknowledgments{\small S.C.\ has been supported by MURST (Cofin2002), and    
H.S.\ by a Marie Curie Fellowship of the   
European Community program "Human Potential" under contract number   
\mbox{HPMF-CT-2000-00951}. S.C., M.S., and H.S.\ acknowledge the hospitality 
of the Max-Planck-Institut f{\"u}r Astrophysik, where part of this work has 
been carried out.  A.W.\ thanks the Institute for Nuclear  
Theory at the University of Washington for its hospitality and the  
Department of Energy for partial support during completion of this  
work.  
We are grateful to A.V.~Sweigart for constructive  
discussions.   
}   

\end{document}